\documentstyle[12pt]{article}
\begin{document}
\baselineskip=24pt
\newcommand{\Lam}{\Lambda_{\scriptscriptstyle {\rm \overline{MS}} }}
\newcommand{\Lamm}[1]{\Lambda {\vspace{-2pt} \scriptscriptstyle ^{^{\rm #1}}
    _{\rm \overline{MS}} }}
\newcommand{\dsp}{\displaystyle}
\newcommand{\dfr}[2]{ \displaystyle\frac{#1}{#2} }
\newcommand{\Lag}{\Lambda \scriptscriptstyle _{ \rm GR} }
\newcommand{\pa}{p\parallel}
\newcommand{\pe}{p\perp}
\newcommand{\paa}{p'\parallel}
\newcommand{\pee}{p'\perp}
\renewcommand{\to}{\rightarrow}
\renewcommand{\baselinestretch}{1.5}
\vspace{-20ex}
\begin{flushright}
\vspace{-3.0ex}
    \it{IC/94/225, AS-ITP-94-24, hep-ph/9408242} \\
\vspace{-2.0mm}
       \it{NEW REVISION June, 1995}\\
%\vspace{-2.0mm}
\vspace{5.0ex}
\end{flushright}

\centerline{\Large\bf On Hadronic Production of the $B_c$ Meson}
\vspace{6.4ex}
\centerline{\large\bf  Chao-Hsi Chang$^{+,*,\dagger}$,
Yu-Qi Chen$^{*,\dagger}$}
\vspace{4mm}
\centerline{\large\bf  Guo-Ping Han$^{\dagger}$ and Hong-Tao
Jiang$^{\dagger}$} \vspace{3.5ex}
\centerline{\sf $^+$International Center for Theoretical Physics, P.O.Box
586, I-34100 Trieste, Italy.}
\centerline{\sf $^*$CCAST (World Laboratory), P.O.Box 8730, Beijing 100080,
China.}
\centerline{\sf $^\dagger$Institute of Theoretical Physics, Academia Sinica,
P.O.Box 2735, Beijing 100080, China.\footnote{Mailing address.}}

\vspace{3ex}
\begin{center}
\begin{minipage}{5in}
\centerline{\large\bf 	Abstract}
\vspace{1.5ex}
\small
{Two of the approaches to the hadronic productions of the
double heavy mesons $B_c$ and $B_c^*$ are investigated.
Comparison in various aspects on the results obtained by the approaches
is made and shown in figures and a table. Some
trial understanding of the approaches themselves and the
achieved results is presented. The results may be used as some
references for discovering the mesons at Tevatron and LHC.}

\vspace{4mm}
\end{minipage}
\end{center}

\newpage

Recently the interest in the $B_c$ meson, one of the double
heavy flavor mesons, is aroused widely due to its
properties.
Similar to the heavy quarkonia, $\eta_c$, $J/\psi$,
and $\eta_b$, $\Upsilon$ etc., it is a double heavy
quark-antiquark bound state, so the QCD inspired potential model
will work well for describing the binding effects of it$^{[1,2]}$;
but different from them, it carries flavors explicitly, so it may
decay by weak interaction only, and as a result, it has a much
longer lifetime ( a typical weak decay
one ) and plentiful decay channels which have sizable
branching ratios $^{[3-6]}$.
Especially, some of its decays can be calculated quite reliably
and may be measurable in the near future $^{[3,4]}$. Thus
the meson $B_c$, in addition to the heavy quarkonia, may be used
to test the QCD inspired potential models and the acting weak
decay mechanisms for relevant heavy flavors further.
Another important reason to make the $B_c$ physics interesting,
is the study of the $B_c$ meson being accessible soon experimentally.
As pointed out by several independent theoretical estimates$^{[3-12]}$,
the cross sections of its production at
certain existent and planned colliders are
sizable and some typical signals may project over the background.

Having all the possible productions of the
double heavy flavored meson reviewed, the authors of
refs.[5,6] have pointed out that the most
suitable ones of high energy
processes to produce sufficient events
of the $B_c$ meson at the existent and planned
facilities, are those at a $Z^0$ boson `factory', such as
LEP-I, and of energetic hadronic collisions at Tevatron and LHC etc.
In ref.[5],
besides a complete calculation on the $B_c$ meson production
at the level of the lowest order of perturbative QCD (pQCD), the
fragmentation functions for $\bar b \rightarrow B_c$ and $\bar b \rightarrow
B_c^*$($S$-wave) were also worked out correctly, while those
for $\bar b \rightarrow
\chi_{(\bar b c)}$ ($P$-wave) in ref.[7]. In fact, it is the first
time to work out the fragmentation functions correctly, because
not all the terms, being the lowest order,
had been taken into account until the authors did.
The fragmentation functions obtained by ref.[5] were confirmed by others
soon$^{[8,11]}$ i.e. the authors of refs.[8,11] recalculated the
same fragmentation functions out in an axial gauge, a different gauge
from adopted in ref.[5], and the factorization of the fragmentation
functions are manipulated more manifestly. Furthermore the evolution
of the fragmentation functions with changing of the fragmentation
energy scale was also considered in refs.[8,11] by solving
the corresponding Altarelli-Parisi equation$^{[13]}$.

Of all the proposed theoretical estimates$^{[5-12]}$,
besides those of pure phenomenological ones
with Monte Carlo simulation such as done in ref.[10],
the adopted approaches for estimating the hadronic productions of
the $B_c$ meson may be divided into two categories, although they
produce results in consistency in order of magnitude (different in values
from each other), and all are based on pQCD.
The first category is to consider the production
in a fragmentation picture i.e. the $B_c$ meson is produced due to
fragmentation of a heavy flavor jet (here $\bar b$ jet mainly)$^{[8]}$.
It is very similar to that of a light meson production
from a jet, but the fragmentation energy
scale is much higher (above that of nonperturbative QCD) that the
fragmentation functions are calculated with pQCD.
According to pQCD, the production cross section:
\begin{equation}\begin{array}{cl}
d\sigma =& \displaystyle \sum_{ijk} \int dx_1 \int dx_2 \int dx_3
F^i_{H_1}(x_1, \mu_F) F^j_{H_2}(x_2, \mu_F) \\[2mm]
 & \cdot d\hat\sigma_{ij\to
kX}(x_1x_2x_3, \mu_F) \cdot D^{B_c}_k(x_3, \mu_F),\\[4mm]
\end{array}\end{equation}
$F^i_{H}(x, \mu_F)$ is the distribution function of the parton $i$ in
the hadron $H$, $d\hat\sigma_{ij\to kX}(\cdots)$ is the cross section
for the relevant jet inclusive production ($i+j \to k+X$) and
$D^{B_c}_k (x, \mu_F)$ is the fragmentation function of $B_c$ from jet
$k$. The formulation here means that the
calculation should be carried out at a typical energy scale $\mu_F$ of
the process. The specific fragmentation
functions $D^{B_c}_k(x, \mu_F), (k=\bar b, c)$ are calculated in the
framework of pQCD. In this approach, it is easy to extend
straightforwardly up to the leading logarithm approximation
(LLA) accuracy level. The second category is
not, as done in the first category,
to factorize the ``subprocess'' $i+j\to B_c+X$ into two
factors further: the `jet production' $i+j\to k+X$
and the `fragmentation' of the $B_c$ meson from the jet $k\to B_c+X$,
but the subprocess is treated as a
whole, and to compute it directly in the framework of pQCD too
i.e. the production cross section:
\begin{equation}
\displaystyle d\sigma=\sum_{ijk} \int dx_1 \int dx_2
F^i_{H_1}(x_1, \mu_F) F^j_{H_2}(x_2, \mu_F) d\hat\sigma_{ij\to
B_cX}(x_1x_2, \mu_F).\\[4mm]
\end{equation}
Although the computations
of the second category are available only up to the lowest
order of pQCD so far, in principle, they may be extended to higher orders
with lengthy and boring calculations. In both of the two approaches
(at the lowest order approximation), the wave function at original of
the $(c \bar b)$ bound state system will occur in the fragmentation
functions and the amplitude of the subprocess $i+j\to B_c+X$ respectively,
whereas the wave function may be obtained from potential
model for the double heavy bound state system
precisely$^{[5,7,8,11]}$. As $B_c^*(1 ^3S_1)$ meson has a
cross section for hadronic productions
bigger than that of $B_c$, and it
will decay to the ground state $B_c$ with a branching ratio almost
100\% in a very `short' time (without decay vertex in experimental detector)
so it contributes the $B_c$ production substantially, thus in the paper
we will discuss $B_c^*$ and $B_c$ together from now on.

In hadronic productions of $B_c(B_c^*)$, the substantial contribution
is from the subprocess of gluon-gluon fusion $g+g \to B_c(B_c^*) +b +\bar c$
but not from quark-antiquark annihilation $q+\bar q \to B_c(B_c^*)+b +\bar c$
at a relatively high energetic colliders
such as Tevatron and LHC etc$^{[6]}$, thus we will restrict
ourselves to discuss the subprocess of gluon-gluon fusion and
to find out the differences attributed to the adopted approaches
from now on in the paper.
For convenience, we will denote the first category as Approach-I,
whereas the second one as Approach-II.

{}From the knowledge of pQCD, Approach-I depends more
on the factorization theorem, whereas Approach-II, being much more
complicated than Approach-I, is of a fixed order complete calculation.
If Approach-I is extended up to the level of
LLA, it may achieve better results for very high energy and high $P_T$
problems. Whereas for some of the other problems with not very high energy
and very high $P_T$, the complete fixed
order approach, Approach-II even at the lowest order
may achieve better ones. We are
interested in examining the two categories of the approaches
quantitatively for the $B_c(B_c^*)$ production not only
due to the the experimental interest, but also
due to the theoretical interest,
because the quantitative results
may offer references for study of $B_c(B_c^*)$
experimentally and understanding the production mechanics theoretically.

The precise differences between the two categories
in the production of the double heavy meson $B_c(B_c^*)$
in $Z^0$ decay, have been given in ref.[5], although there
the LLA corrections for Approach-I were not considered and
the results for the rate of $B_c(B_c^*)$ production were underestimated
owing to a smaller QCD coupling ($\alpha_s(Q^2)$ $Q^2=m_{Z^0}^2$) being
adopted. Since the relevant $\bar b$-jet produced in $Z^0$
decay is very energetic ($E_b=m_{Z^0}/2$ at C.M.S.) and
the process is comparatively simple, thus to find the correspondece
of the approaches is simple, the difference in values between the
approaches in partial width is less than $20\%$. However, in the
hadronic productions the colliding energy of the subprocess
varies in a wide region and may be quite `low' at Tevatron, even at LHC,
and the subprocess itselfis much more complicated.
The energies of the subprocess in the hadronic collision at Tevantron,
even at LHC, in most chances are much
smaller than that in $Z^0$ decay, and
we will return this point more precisely later on. Furthermore
the subprocess is much more complicated
than that of $Z^0$ decay: there are 3 diagrams instead of one of $Z^0$ decay
for Approach-I, and there are 36 diagrams instead of 2 for Approach-II.
Concerning the hadronic $B_c(B_c^*)$ production, the masses
of the heavy quarks inside the meson $B_c(B_c^*)$ play
the role as the proper energy scale in the production, and it
is much greater than $\Lambda_{QCD}$ so pQCD calculations are
always applicable, no matter how big $P_T$ the heavy flavor
(here the $B_c(B_c^*)$ meson) carries, that is very different from
light flavor productions. Generally to know the accurate contributions from
the low $P_T$ components quantitatively, which is expected to have substantial
different results for the approaches, is interesting, because
it is necessary for writing an event generator which may produce
reliable low $P_T$ events. To see how low $P_T$ events
could be well detected by the concrete detector is interesting for the
experiments on the concerned subject(s) i.e. $B_c(B_c^*)$
mesons for present problem, whereas without the reliable event
generator, the aim could not be reached at.

The differences between the two categories of the approaches, in fact,
may be attributed how to deal with the subprocess of Approach-II.
In Approach-II we deal with it as a whole i.e. a complete pQCD calculation
on the process, though only the lowest order one is available so far;
whereas in Approach-I it is treated to produce heavy quark jets
first and then to fragment a meson $B_c(B_c^*)$ from one of the produced heavy
jets, thus as known from the proof of the pQCD factorization theorem,
Approach-I is not as good as Approach-II even doubtable, if the jet
responsible for the fragmentation of a $B_c(B_c^*)$ meson, is not very
energetic.

To understand the differences of the approaches, let us analysize the
subprocess, $g+g\rightarrow B_c(B_c^*)+b+\bar{c}$, carefully.
According to Approach-II, there are 36 Feymann diagrams
responsible for it. Some of the typical ones are collected in Fig.1(a,b).
It is easy to realize that the diagrams
(the amplitude) may be divided (splitted) into 5 independent subgroups
(terms) according to their color structure, and each of them alone
is guage invariant$^{[6]}$. It is too long
to write down here the total amplitude of the subprocess
explicitly, however, we may write its color structure out
explicitly in a short formulation, and with it
we will be able to find out some correspondence and difference
of the two approaches.
In general, the formulation for the amplitude:
\begin{equation}
A(a,b,i,j)=\dsp\sum_{\alpha=1}^{6} C^{ab}_{\alpha ij}M_{\alpha}(
\epsilon_1,\epsilon_2,s_1,s_2).\\[4mm]
\end{equation}
Here each of the color factors
$C^{ab}_{\alpha ij} (\alpha=1,2,\cdots,6)$
is a product of
the Gell-Mann matrices:
\begin{equation}
\begin{array}{lll}
\displaystyle C_{1\;ij}^{ab} &=\displaystyle (\lambda^c \cdot \lambda^c
\cdot \lambda^a \cdot \lambda^b)_{ij}
&=\displaystyle \dfr{N^2-1}{N}(\lambda^a \cdot \lambda^b)_{ij}\; ;\\[2mm]
\displaystyle C_{2\;ij}^{ab} &=\displaystyle (\lambda^c \cdot \lambda^c
\cdot \lambda^b \cdot \lambda^a)_{ij}
&=\displaystyle \dfr{N^2-1}{N}(\lambda^b \cdot \lambda^a)_{ij}\; ;\\[2mm]
\displaystyle C_{3\;ij}^{ab} &=\displaystyle (\lambda^c \cdot \lambda^a \cdot
\lambda^c \cdot \lambda^b)_{ij}
&=\displaystyle \dfr{-1}{N}(\lambda^a \cdot \lambda^b)_{ij}\; ;\\[2mm]
\displaystyle C_{4\;ij}^{ab} &=\displaystyle (\lambda^c \cdot \lambda^b \cdot
\lambda^c \cdot \lambda^a)_{ij}
&=\displaystyle \dfr{-1}{N}(\lambda^b \cdot \lambda^a)_{ij}\; ;\\[2mm]
\displaystyle C_{5\;ij}^{ab} &=\displaystyle (\lambda^c \cdot \lambda^a \cdot
\lambda^b \cdot \lambda^c)_{ij}
&=\displaystyle \delta_{ij}{\rm tr} (\lambda^a \cdot \lambda^b)-\dfr{1}{N}
(\lambda^a \lambda^b)_{ij}\; ;\\[2mm]
\displaystyle C_{6\;ij}^{ab} &=\displaystyle (\lambda^c \cdot \lambda^b \cdot
\lambda^a \cdot \lambda^c)_{ij} &=\displaystyle \delta_{ij}{\rm tr}
(\lambda^a \cdot \lambda^b)-\dfr{1}{N} (\lambda^b
\cdot \lambda^a)_{ij}.\\[4mm]
\end{array}\end{equation}
However, we should note here that not all these color factors are independent,
because there exists a relation among them, that is
\begin{equation}
\displaystyle C_{3\;ij}^{ab} - C_{5\;ij}^{ab} = C_{4\;ij}^{ab} -
C_{6\;ij}^{ab}.\\[4mm]
\end{equation}
Therefore only 5 color
factors are independent, and we may choose them as:
\begin{equation}
\begin{array}{l}
C_{m\;ij}^{'ab}=  C_{m\;ij}^{ab} \;\;\;(\; {\rm when }\; \;m=1,\cdots,
4 \;);\\[4mm]
C_{5\;ij}^{'ab} = C_{3\;ij}^{ab} - C_{5\;ij}^{ab}. \\[4mm]
\end{array}
\end{equation}
Thus the amplitude may be rewritten as:
\begin{equation}
A(a,b,i,j)=\dsp\sum_{k=1}^{5} C'^{ab}_{k\;ij}M'_{k}(
\epsilon_1,\epsilon_2,s_1,s_2).
\vspace{4mm}
\end{equation}

Being independent, the coefficients of the color factor $C^{'ab}_{k\;ij}$,
the sub-amplitudes $M'_{k}$ ( $k=1,2,\cdots,5$ )
are individually gauge invariant, thus each of them may acquire certain
meaning.
Owing to the fact that each of the amplitudes $M'_{k}$
is related to certain Feynman diagrams of the 36 precisely,
the explicit formulas of $M'_{k}$ ( $k=1,2,\cdots,5$ )
may be written down directly, based on the rules
of the duel amplitude method$^{[8]}$. Therefore one may find out
the correspondences and difference between Approach-I and Approach-II.
Since the Approach-I is of a fragmentation of a $\bar b$-quark jet
in the $B_c(B_c^*)$ production (the fragmentation of a $c$-jet contributes too,
but it is much less important than that of a $\bar b$-jet),
with the decomposition eqs.(4-7) and according to the color structure
to trace back to the diagrams, one may find the correspondence:
the sub-amplitudes $C'^{ab}_{k\;ij}M'_k$ $(k=1,2)$ is to correspond
to some of Approach-I's amplitudes, whereas those of
$C'^{ab}_{k\;ij}M'_k$ $(k=3,4,5)$ cannot find
any correspondence in Approach-I. The fact of the correspondences
is easy to be understood by means of the Feynmann diagrams of the
two approaches: One may find that the diagrams
such as Fig.1(a) which contribute to the sub-amplitudes
$C'^{ab}_{k\;ij}M'_k$ $(k=1,2)$ substantially
in the sense of the color structure, could be understood
as if two jets were produced and the fragmentation of $B_c(B_c^*)$ meson
was followed, whereas for the diagrams such as Fig.1(b),
which contribute to the sub-amplitudes $C'^{ab}_{k\;ij}M'_k$ $(k=3,4,5)$
substantially, there is no similar correspondence at all in
the above sense to Approach-I.
Therefore we expect the results achieved by the two approaches
being different, so a thorough investigation
of the approaches quantitatively, even though numerically, is interesting.
We will devote this paper to the investigation\footnote{ During
the period of revising the paper, several papers$^{[19,20]}$ come out
and certain disagreements on the calculations
are presented, thus to clarify the situation is also necessary.}, i.e.
to compare the $B_c(B_c^*)$ hadronic productions of the approaches
quantitatively in various aspects. We will plot
the numerical results of each observable, obtained by the two
approaches into one figure together, different figures show different
aspects of the approaches, and finally we will try to
reach at some conclusions.

First of all, we calculate the total cross sections
of the $B_c(B_c^*)$ productions by the two approaches
at various hadronic colliders i.e. for various C.M.S's energies
of the colliding hadrons, but only the lowest order
for the subprocess $g+g\to B_c(B_c^*)+b+\bar c$, is concerned.
The obtained total cross sections are put into Tab.1. We should
note here that throughout the paper without special statement,
the following manipulations and parameters are taken.
When calculating the productions of $p-p$ and $p-\bar p$ collisions,
only gluon-gluon fusion mechanism is considered due to its
domination over the others such as quark-antiquark etc$^{[6]}$;
the CTEQ3 structure functions with $\Lambda^{QCD}_{\bar{MS}}(n_f=4)=0.239GeV$
(corresponding $\alpha_s(m_Z^2)=0,112$)$^{[14]}$
and $\alpha_s(Q^2)$ with an energy scale $Q^2=\bar{s}/4$
($\bar{s}$ is the c.m. energy squared
of the subprocess) are adopted. As for the masses,
the values $m_c=1.5GeV$, $m_b=4.9GeV$ and $M_{B_c(B_c^*)}=6.4GeV ^{[2]}$
are taken. Furthermore in the calculations the wave functions of $B_c$ and
$B_c^*$ at origin are obtained obtained from potential model and
the difference of wave functions, as the masses, for $B_c$ and $B_C^*$
is ignored here for the `lowest order calculation'.
In order to compare with those adopted in literature easy,
what we adopt it here is in decay constant formulation: $f_{B_c}\simeq
480MeV$ (under the convention $f_\pi=132MeV$).

Note here that in the table when calculating the subprocess
$g+g\to B_c(B_c^*)+b+\bar c$ at $\sqrt {\bar s}=20, 30, 60 GeV$,
a constant of strong coupling $\alpha_s=0.2$ is taken, and
when the row is denoted with a `$*$' (`$**$'), the results are indicated
to have a cut for small $P_T\leq 5 GeV$ ($P_T\leq 10 GeV$). Up to the concerned
order of pQCD, the uncertainties here come only from the choices
of the values of $m_c$, $m_b$, $M_{B_c(B_c^*)}$, $\alpha_s$
and $f_{B_c}$.

\begin{center}
{\bf TABLE I. The total cross sections for the productions of
the $B_c$ meson and its excited state $B_c^*$ obtained by the two approaches
(in unit $nb$).\\
\vspace{4mm}
\begin{footnotesize}
\begin{tabular}{c|cc|cc} \hline\hline
Collision & \multicolumn{2}{c|}{Approach-I}
&\multicolumn{2}{c}{Approach-II}\\[.018in]
\hline
& $B_c\;(1^1S_0)$ & $B_c^\ast\;(1^3S_1)$  & $B_c\;(1^1S_0)$
& $B_c^\ast\;(1^3S_1)$ \\[.018in]
\hline
$p\bar{p} (\sqrt S=1.8TeV)$ & $0.747(4)$ & $1.23(1)$ & $0.850(8)$ & $2.07(2)$
\\[.018in]
\hline
$p\bar{p} (\sqrt S=1.8TeV, *)$ & $0.229(2)$ & $0.389(3)$ & $0.259(4)$
& $0.646(6)$ \\[.018in]
\hline
$p\bar{p} (\sqrt S=1.8TeV, **)$ & $0.0331(9)$ & $0.0570(6)$ & $0.0373(1)$
& $0.0894(3)$ \\[.018in]
\hline
$pp (\sqrt S=14TeV)$ & $8.63(5)$ & $14.0(1)$ & $10.6(1)$ & $26.4(3)$ \\[.018in]
\hline
$pp (\sqrt S=14TeV, *)$ & $3.07(3)$ & $5.11(4)$ & $3.71(6)$ & $9.43(9)$
\\[.018in]
\hline
$pp (\sqrt S=14TeV, **)$ & $0.584(7)$ & $0.986(10)$ & $0.698(1)$ & $1.69(4)$
\\[.018in]
\hline
$gg (\sqrt {\bar s}=20GeV)$ & $0.704(5)\cdot 10^{-2}$ & $0.118(1)\cdot 10^{-1}$
& $0.661(7)\cdot 10^{-2}$ & $0.160(2)\cdot 10^{-1}$ \\[.018in]
\hline
$gg (\sqrt {\bar s}=30GeV)$ & $0.678(8)\cdot 10^{-2}$ & $0.103(1)\cdot 10^{-1}$
& $0.949(8)\cdot 10^{-2}$ & $0.244(3)\cdot 10^{-1}$ \\[.018in]
\hline
$gg (\sqrt {\bar s}=60GeV)$ & $0.321(7)\cdot 10^{-2}$ & $0.456(9)\cdot 10^{-2}$
& $0.782(9)\cdot 10^{-2}$ & $0.203(3)\cdot 10^{-1}$ \\[.018in]
\hline\hline
\end{tabular}
\end{footnotesize}
}
\end{center}

\vspace{4mm}

The $P_T$ dependence of the productions at various colliders
Tevatron and LHC is interesting experimentally, thus we have calculated
it and plotted the results in Fig.2. In the calculations,
the low $P_T$ component contribution has been taken
into account too, though for Approach-I the computation is
problematic. It is because the production closing to the threshold
(where $P_T$ cannot be big)
needs special consideration and corrections in Approach-I, but here
we merely make the `approximation': the $P_T$ of $B_c(B_c^*)$ being fixed
in the direction of the produced heavy quark jet, in fact,
it is not a good approximation when the `fragmentation'
is very close to the threshold of the $B_c(B_c^*)$ meson production, thus
the low $P_T$ component contribution as shown in Fig.2 is not so
well estimated for Approach-I. From the figure one may see that
for the $B_c$ production, the difference between the two approaches
is not sizable but for the $B_c^*$ production it is quite great (about
a factor two even greater) and, general speaking, as $P_T$ is going
high the production cross sections predicted by the two
approaches are approaching to equal (for $B_c^*$, up to $P_T=20 GeV$
they are still different).

In order to have an outline about the gluon-gluon subprocess in hadronic
collisions, in Fig.3, we present the production cross sections at Tevatron
and LHC versus the collision energy $\bar s$ of the glouns inside the
collision hadrons. As the small $P_T$ component of the productions is
not able to measure, we have imposed a cut for those of small
$P_T (\leq 5 GeV)$ here. From the figure, one may see the
cross sections drop in a logarithm scale versus $\bar s$ increasing.
In fact, if we had not imposed the cut for small $P_T$, the cross sections
would have a ``peak'' around $20 GeV$ ( not very far from the threshold
of the subprocess $\sqrt {\bar s} \sim 12.8 GeV$ ), and then would drop
\footnote{To shorten the paper and to present the more useful results,
we would not present the curves without $P_T$ cut here, although we have
them.}. One may see that when $\bar s$ reaches at $80 GeV$, the
cross sections have dropped down at least one more orders of the magnitude
already. Thus one may conclude that in the hadronic collisions the dominant
contribution to the $B_c$ and $B_c^*$ meson productions is not
from very high energetic gluon fusion but from relatively low energy,
that is the great difference from that in $Z^0$ decay as we emphasized
earlier in the paper. For $B_c^*$ production,
the cross sections obtained by Approach-II are greater than
those by Approach-I at various energies with a factor $5$ or greater,
but for $B_c$ production, the difference caused by the two
approaches is within a factor $2$, less than that for
$B_c^*$ production.

We should note here that besides the cut for small $P_T$ being imposed
and the coupling constant $\alpha_s$ being running,
the cross sections in Fig.3 are different in meaning
from that of the gluon-gluon fusion for precise $\sqrt{\bar s}$
in Tab.1, as the later is merely of gluon-gluon fusion
but the former has the structure functions of the collision
hadrons convoluted into.

All the resultant cross sections of the hadronic
productions are achieved always
by a convolution of the cross section of
the relevant subprocess and a common factor,
the structure functions of the incoming hadrons
of the collisions. In order to highlight the
differences of the two approaches, we have also calculated the
subprocess cross sections as if the subprocess is an independent one,
i.e. the cross sections of the gluon-gluon fusion at
various precise energies. The total cross sections have been put in Tab.1
already, but the transverse momentum $P_T$ and rapidity $Y$ distributions
at various C.M.S's energies are presented in Figs.4:
in Fig.4(a,b) for $\sqrt{\bar s}=30~GeV$, Fig.4(c,d) for
$\sqrt{\bar s}=60~GeV$ respectively. In these calculations, we have
taken $\alpha_s=0.2$, a constant, as emphasized above.
One may see the fact very clearly that, as expected, the values obtained
by Approach-II at small $P_T$ and small $Y$ are always greater than those
obtained by Approach-I, whereas the values are approaching close when $P_T$
or $Y$ increases.

In summary, the two approaches cause some substantial differences
in total cross sections and the $P_T$ distribution etc. indeed,
especially at low $P_T$. Approach-II should be
suitable at low $P_T$, even at low $P_T$ and low $Y$ both.
The heavy masses of the quarks inside the meson play the role
to offer the least and proper energy scale in the concerned productions
and to guarantee pQCD being applicable always. Namely it is the heavy
quark masses being the least proper energy scale, instead of
$\Lambda_{QCD}$, that appear in the formulas (appear in $\alpha_s$,
the coupling constant for the lowest order calculations
and in the logarithmically large terms if higher order calculations
are carried out).
Furthermore, the most important productions of the $B_c(B_c^*)$
meson are shown in Fig.3 not due to very energetic parton collisions,
and from Fig.2 one may also see that the $P_T$ cannot be great in
the interesting processes concerned in the paper,
thus in order to collect as many as possible events
of the mesons $B_c$ and $B_c^*$ so as to discover them
and to study their properties, one could not estimate
the low $P_T$ components of the productions
too rughly from very beginning and should try to
have a good one as one can. For this purpose, it is sure that
Approach-II is good, and the logarithmical
terms to the heavy quark masses need not
to worry about too much, as the terms appear
from high order calculations and become important at very large $P_T$
only$^{[16]}$. Approach-I is better than Approach-II for
estimating the productions at very large $P_T$
if the former taking into account the large logarithmical terms
to the heavy quark masses by LLA but the later not (as the present case).
Whereas for the concerning hadronic productions at Tevatron, even at LHC,
the former's advantages have not ``matured'' yet because of the same reason
as pointed out above: $P_T$ cannot be very great at the concerned
processes.

We think that the defferences of the obtained results by the two approaches
may be understood by the fact that Approach-II involves more
mechanisms than those Approach-I does, as argued in terms of eqs.
(3-7) early. Recently the authors of ref.[17] also recognized that
certain higher order gluon fragmentation besides the fragmentation of
a heavy quark may contribute to $B_c$ production substantially.

According to the experiences of heavy quark
productions in hadronic collisions and the theoretical loop
calculations, we know that a full perturbative QCD calculation
up to one loop level may achieve quite high accuracy$^{[18]}$,
thus a higher order full perturbative QCD calculation on
the hadronic production of the double heavy flavor meson
$B_c(B_c^*)$ under Approach-II will be very interesting surely$^{[16]}$.

In the procedure of revising the paper, several calculations$^{[19-21]}$
on the same problems appear. The authors of ref.[20] have found that
in their earlier version they had omitted a color factor $1/\sqrt 3$
in amplitude, and when having the mistake corrected they have
found a nice agreement between theirs and those
of ref.[6]. The authors of ref.[19] have
investigated various calculations quite systematically, therefore
we have checked the numerical results for the subprocess $g+g\to
B_c(B_c^*)+b+\bar c$
at $\sqrt {\bar s}= 20, 40 GeV$, by means of our program but
having their parameters. As a result, we have found that our results and theirs
are consistent with each other exactly within the Monte Calor errors\footnote
{In the early version of this paper, there was a mistake in the programs
for the numerical calculations, thus there were some disagreements between
our earlier results and those of ref.[19].}, that in fact is a confirmation
for our programs and those of ref.[19]. Only for updating, the new version
of the structure functions CTEQ3$^{[14]}$ and the parameters
appearing in the calculations as quoted above
(e.g. the energy of HLC $\sqrt S=14 GeV$ etc), which are
slight different from those of ref.[19] even ref.[6], have been adopted
here so that the numerical results for the total cross sections and the other
obsevables involving two structure functions of the two colliding
hadrons are reasonably different from those of refs.[6,19,20] a little.
However there are some of disagreements between our results and those of
ref.[21].

\vspace{4ex}
{\Large\bf{ Acknowledgements}}

The work was supported in part by the National Natural Science Foundation
of China and the Grant LWTZ-1298 of Chinese Academy of Science.
The authors would like to thank Prof. Yuan-Ben Dai for useful
discussions. One of the author (Chao-Hsi Chang) would like to thank
Profs. A. Salam and S. Randjbar-Daemi, the International Atomic
Energy Agency and UNESCO for hospitality during his stay at ICTP.
Finally the authors would like to thank Dr. H.L. Lai for offering
the numerical program of the structure functions and the referees for useful
suggestions.

\newpage

\centerline{\large\bf{ Figure Captions}}

\vspace{1ex}

\begin{enumerate}

\item Some Typical Feymann Diagrams for the Subprocess $g+g\to
B_c(B_c^*)+b+\bar c$.
(Here the spots in the figure denote the bound state $B_c(B_c^*)$. In
fact, there are 36 diagrams in total, but here we plot some of them only for
illustrating the existence of various mechanisms involved in the process.)

\begin{enumerate}

\item The Feymann Diagrams Correspond to that for Fragmentation
mechanism mainly.

\item The Feymann Diagrams Correspond to those for Some else besides the
Fragmentation Mechanism mainly.

\end{enumerate}

\item The the Transverse Momentum $P_T$ Distribution for $B_c(B_c^*)$
Productions at Tevatron and LHC for Approach-I and Approach-II.

\begin{enumerate}

\item The Differential Cross Sections of the $B_c(B_c^*)$ Productions
versus the the Transverse Momentum $P_T$ at Tevatron. The solid line
($A~V$) and
the dashed line ($C~V$) are those of the $B_c^*$ productions obtained by
Approach-I and Approach-II respectively; The dashed-dotted line ($A~P$) and
the dotted line ($C~P$) are those of the $B_c$ productions obtained by
Approach-I and Approach-II respectively.

\item The Differential Cross Sections of the $B_c(B_c^*)$ Productions
versus the the Transverse Momentum $P_T$ at LHC. The notation
of the lines is the same as Fig.2a.

\end{enumerate}

\item The Differential Cross Sections of the $B_c(B_c^*)$ Productions
versus the C.M. Energies of the Colliding Gluon-Gluon at the Colliders
for Approach-I and Approach-II.

\begin{enumerate}

\item The Differential Cross Sections of the $B_c(B_c^*)$ Production
versus the C.M. Energies of the Colliding Gluon-Gluon at Tevatron.
The notation of the lines is the same in Fig.2a.

\item The Differential Cross Sections of the $B_c(B_c^*)$ Production
versus the C.M. Energies of the Colliding Gluon-Gluon at LHC. The
notation of the lines is the same in Fig.2a.

\end{enumerate}

\item The Distributions of the Transverse Momentum $P_T$ and the Rapidity $Y$
for $B_c(B_c^*)$ Productions at Precise Energies ($\sqrt {\bar s}$)
of the Gluon-Gluon Collisions Respectively.

\begin{enumerate}

\item The Differential Cross Sections of the $B_c(B_c^*)$ Productions
versus the Transverse Momentum $P_T$  at $\sqrt {\bar s }= 30 GeV$.
The notation of the lines is the same in Fig.2a.

\item The Differential Cross Sections of the $B_c(B_c^*)$ Productions
versus the Rapidity $Y$  at $\sqrt {\bar s} = 30 GeV$.
The notation of the lines is the same in Fig.2a.

%\end{enumerate}

%\item The Differential Cross Sections of the $B_c$ and $B_c^*$ Productions
%versus the Produced Transverse Momentum $P_T$ and the Rapidity $Y$
%Respectively at $\sqrt {\bar s} = 30 GeV$.

%\begin{enumerate}

%\item The Differential Cross Sections of the $B_c$ and $B_c^*$ Productions
%versus the Produced Transverse Momentum $P_T$ at $\sqrt {\bar s} = 30
%GeV$. The notation of the lines is the same in Fig.2(a).

%\item The Differential Cross Sections of the $B_c$ and $B_c^*$ Productions
%versus the Rapidity $Y$ at $\sqrt {\bar s} = 30 GeV$. The notation of the
%%lines
%is the same in Fig.2(a).

%\end{enumerate}

%\item The Differential Cross Sections of the $B_c$ and $B_c^*$ Productions
%versus the Produced Transverse Momentum $P_T$ and the Rapidity $Y$
%Respectively at $\sqrt {\bar s }= 60 GeV$.

%\begin{enumerate}

\item The Differential Cross Sections of the $B_c(B_c^*)$ Productions
versus the Transverse Momentum $P_T$ at $\sqrt {\bar s} = 60 GeV$.
The notation of the lines is the same in Fig.2a.

\item The Differential Cross Sections of the $B_c(B_c^*)$ Productions
versus the Rapidity $Y$ at $\sqrt {\bar s }= 60 GeV$. The notation of the lines
is the same in Fig.2a.

\end{enumerate}

\end{enumerate}

\end{document}